\documentclass[twocolumn,prd,aps,superscriptaddress,showpacs,amsmath,amssymb]{revtex4-2}
\usepackage{graphicx}
\usepackage{amssymb}
\usepackage{amsmath}
\usepackage{epsfig}
\usepackage{color}
\usepackage{mathtools}
\usepackage[colorlinks,linkcolor=blue,anchorcolor=blue,citecolor=blue,urlcolor=blue]{hyperref}
\usepackage{physics}
\usepackage{bm}
\usepackage{diagbox}
\usepackage{array}

\def \M {\mathsf{M}}
\def \MM {\mathcal{M}}
\def \L {\mathsf{L}}
\def \LL {\mathcal{L}}
\def \P {\mathsf{P}}
\def \PP {\mathcal{P}}
\def \T {\mathcal{T}}

\def \Z {\mathbb{Z}}
\newcommand{\PreserveBackslash}[1]{\let\temp=\\#1\let\\=\temp}
\newcolumntype{C}[1]{>{\PreserveBackslash\centering}p{#1}}

\begin{document}

\title{Periodic Clifford symmetry algebras on flux lattices}

\author{Yue-Xin Huang}
\affiliation{Research Laboratory for Quantum Materials, Singapore University of Technology and Design, Singapore 487372, Singapore}
\author{Z. Y. Chen}
\affiliation{National Laboratory of Solid State Microstructures and Department of Physics, Nanjing University, Nanjing 210093, China}
\author{Xiaolong Feng}
\affiliation{Research Laboratory for Quantum Materials, Singapore University of Technology and Design, Singapore 487372, Singapore}
\author{Shengyuan A. Yang}
\affiliation{Research Laboratory for Quantum Materials, Singapore University of Technology and Design, Singapore 487372, Singapore}
\affiliation{Center for Quantum Transport and Thermal Energy Science,
School of Physics and Technology, Nanjing Normal University, Nanjing 210023, China}
\author{Y. X. Zhao}
\email{zhaoyx@nju.edu.cn}
\affiliation{National Laboratory of Solid State Microstructures and Department of Physics, Nanjing University, Nanjing 210093, China}
\affiliation{Collaborative Innovation Center of Advanced Microstructures, Nanjing University, Nanjing 210093, China}

\begin{abstract}
    Real Clifford algebras play a fundamental role in the eight real Altland-Zirnbauer symmetry classes and the classification tables of topological phases. Here, we present another elegant realization of real Clifford algebras in the $d$-dimensional spinless rectangular lattices with $\pi$ flux per plaquette. Due to the $T$-invariant flux configuration, real Clifford algebras are realized as projective symmetry algebras of lattice symmetries. Remarkably, $d$ mod $8$ exactly corresponds to the eight Morita equivalence classes of real Clifford algebras with eightfold Bott periodicity, resembling the eight real Altland-Zirnbauer classes.
    The representation theory of Clifford algebras determines the degree of degeneracy of band structures, both at generic $k$ points and at high-symmetry points of the Brillouin zone. Particularly, we demonstrate that the large degeneracy at high-symmetry points offers a rich resource for forming novel topological states by various dimerization patterns, including a $3$D higher-order semimetal state with double-charged bulk nodal loops and hinge modes, a $4$D nodal surface semimetal with $3$D surface solid-ball zero modes, and $4$D M\"{o}bius topological insulators with a eightfold surface nodal point or a fourfold surface nodal ring. Our theory can be experimentally realized in artificial crystals by their engineerable $\mathbb{Z}_2$ gauge fields and capability to simulate higher dimensional systems.
\end{abstract}

\maketitle

\section{Introduction}

Clifford algebra is a mathematical structure that has found applications in diverse subjects such as modern geometry~\cite{atiyah1964clifford}, quantum field theory, supergravity~\cite{polchinski1998string}, and digital image processing~\cite{rodriguez2008action}. Its subset, the real Clifford algebras (RCAs), recently came into the limelight in
condensed matter physics, when they were revealed to underlie the classification theory of topological band structures~\cite{schnyder_classification_2008,kitaev_periodic_2009,zhao_PRL_2013,chiu_classification_2016,zhao_PRL_2016a}.
Considering the time-reversal ($T$) and particle-hole ($C$) symmetries, there are ten symmetry classes, known as the Altland-Zirnbauer classes~\cite{AZ_Classes},
eight of which containing at least $T$ or $C$ symmetry are real and are described by the RCAs~\cite{kitaev_periodic_2009}.
The theory of RCA helps to establish the topological classification table~\cite{chiu_classification_2016}, which exhibits an intriguing eightfold periodicity along spatial dimensions, rooted in the famous Bott periodicity of RCAs under the Morita equivalence~\cite{atiyah1964clifford,morita_duality_1958}.

In this Letter, we present another elegant realization of RCAs by the projective symmetry algebras of simple rectangular lattices with $\Z_2$ gauge fluxes. The $\Z_2$ gauge fields, namely allowing the (real) hopping amplitudes to take positive or negative signs, can be readily engineered in various artificial crystals, such as acoustic/photonic crystals~\cite{lu_topological_2014,yang_topological_2015,mittal_photonic_2019,xue_observation_2020}, cold atoms in optical lattices~\cite{jaksch_creation_2003,aidelsburger_experimental_2011,aidelsburger_realization_2013}, electric-circuit arrays~\cite{imhof_topolectrical-circuit_2018,yu_4d_2020}, and mechanical systems~\cite{huber_topological_2016,prodan_topological_2009}. The importance of projective symmetry algebras under $\Z_2$ gauge fields has recently been emphasized~\cite{zhao_mathbbz_2-projective_2020,zhao_switching_2021,shao_gauge-field_2021}, with fascinating consequences demonstrated in experiment~\cite{xue_projectively_2021,li_acoustic_2021}. Meanwhile, artificial crystals also enable simulations of higher-dimensional states by synthesized dimensions, e.g., $4$D quantum Hall states has been demonstrated experimentally~\cite{price_four-dimensional_2015,lohse_exploring_2018,wang_circuit_2020}.

Motivated by these recent experimental progress, we consider spinless rectangular lattices in $d$ dimensions, with $\pi$-flux for each plaquette. Their projective symmetry algebras are generated by lattice translations, mirror reflections, and $T$.
We show that RCAs naturally arise as projective algebras of these symmetries in momentum space. It follows that the RCA theory elegantly captures the symmetry characters of the band structures. (i) For a generic $k$-point in the Brillouin zone (BZ), $d=1,2,\cdots, 8$ exhaust all eight Morita equivalence classes of RCAs, resembling the eight real Altland-Zirnbauer classes. Interestingly, there again emerges an eightfold periodicity: If $d_1=d_2\mod 8$, they correspond to the same Morita class, such that all dimensions $>8$ can be reduced to $d=1,\cdots,8$. (ii) For any high-symmetry point $K^{s,r}$ in BZ, with $s$ ($r$) the number of $\pi$ ($0$) components in its momentum-space coordinate, the little algebra is also characterized by a RCA determined by $r=d-s$.
We show that the irreducible representations (IRREPs) of RCAs determine the band structure degeneracy. Particularly,
large degeneracies are dictated at certain high-symmetry points, which are resources transformable to rich novel topological phases under proper symmetry breaking. We demonstrate concrete examples with $3$D higher-order nodal-loop semimetals, $4$D nodal-surface semimetals, and $4$D M\"{o}bius insulators, which can be realized using artificial crystals.

\section{Projective symmetry algebra}

A RCA $C^{p,q}$ is generated by $(p+q)$ generators $e_\mu$ ($\mu=1,\cdots, p+q$) which satisfy
\begin{equation}
  \{e_\mu, e_\nu\}=\eta_{\mu\nu}1,
\end{equation}
where  $\eta_{\mu\nu}$ is the diagonal matrix with first $p$ diagonal elements being $-1$ and the other being $1$.
Below, we show that this algebra is naturally realized by the symmetries of a rectangular lattice in a projective way.

Consider a rectangular lattice in $d$ dimensions. Its space group is generated by lattice translations $L_a$ and mirror reflections $M_a$ with $a=1,2,\cdots, d$. Here, $M_a$ reverses the $a$th coordinate $x_a$, and $L_a$ translates $x_a$ by a lattice constant $c_a$. Note that there are two types of mirrors: those coinciding with a lattice-site plane and those between two neighboring planes. To be specific, we choose the latter type, and the other mirrors can be generated by combinations of  $L_a$ and $M_a$. The inversion symmetry $P$ is also generated as $P=\prod_a M_a$.

\begin{figure}
    \centering
    \includegraphics[width=0.4\textwidth]{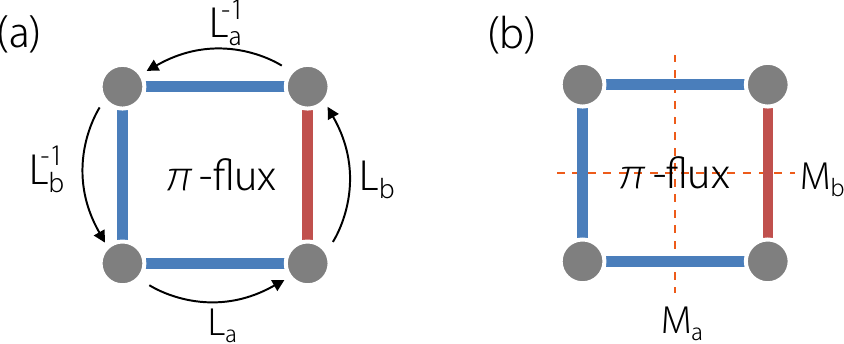}
    \caption{(a) $\L_b^{-1}\L_a^{-1}\L_b\L_a$ moves a particle around one plaquette which encloses a $\pi$ flux.
        (b) The successive mirror operations
        $\M_b\M_a\M_b\M_a$ also encloses an odd number of $\pi$ flux. Here, the red (blue) bonds have negative (positive) hopping amplitudes. }
    \label{fig-LMop}
\end{figure}
Under gauge fields, the symmetries will be projectively represented.  With a $\pi$-flux for each rectangular plaquette, the projective translations $\L_a$ become anti-commutative with each other~\cite{zhao_mathbbz_2-projective_2020},
\begin{equation}\label{Trans_Alge}
\{\L_a,\L_b\}=0\qquad (a\neq b).
\end{equation}
This relation can be readily understood from the Aharonov-Bohm effect, since $\L_b^{-1}\L_a^{-1}\L_b \L_a=-1$ just corresponds to the phase $\pi$ accumulated from circulating a plaquette [Fig.~\ref{fig-LMop}(a)]. Meanwhile,
the projectively represented mirrors $\M_a$ satisfy a RCA:
\begin{equation}\label{M_Cliff}
		\{\M_a,\M_b\}=2\delta_{ab},
\end{equation}
because for any lattice site $\bm{x}$, the path $\M_b\M_a\M_b\M_a(\bm{x})$ encloses an odd number of plaquettes when $a\ne b$ [Fig.~\ref{fig-LMop}(b)].
Similarly, one can show that
\begin{equation}\label{M_L_Alg}
	\M_a \L_a \M_a=\L^{-1}_a,\quad \{\M_a,\L_b\}=0
\end{equation}
with $a\ne b$. Below, we shall make use of the inversion symmetry $\P=\prod_a \M_a$, so let's also lay out its algebraic relations
\begin{equation}\label{P_L_Alge}
	\P \L_a\P=(-1)^{\frac{1}{2}(d+2)(d-1)}\L_a^{-1},\quad \P^2=(-1)^{\frac{1}{2}d(d-1)}.
\end{equation}

It must be stressed that the $\pi$ fluxes preserve the $T$ symmetry. In addition,
since all hopping amplitudes are real, $\M_a$ and $\L_a$ are real operators, and therefore must commute with  $T=\mathcal{K}$ with $\mathcal{K}$ the complex conjugation:
\begin{equation}\label{T_ML_Alg}
	[\M_i,T]=0,\quad [\L_i,T]=0.
\end{equation}

To analyze the constraint of the projective symmetry algebra on band structures, we need to represent it in momentum space.
To this end, we first have to specify the BZ by choosing an Abelian subalgebra of the projective algebra \eqref{Trans_Alge} of translations.
Here, we choose the one generated by $\L_a^2$ with $a=1,\cdots, d$. It is natural for two reasons. First, it treats all directions on the equal footing. Second, it is a group algebra normal to the projective symmetry algebra, i.e., closed under conjugation by any generator. Then, we can represent the free Abelian group generated by $\L_a^2$ as
\begin{equation}
	\LL_a^2=e^{ik_a}.
\end{equation}
Hereafter, we use the calligraphy font to denote representation in momentum space. Then, $\T=\mathcal{K}\hat{I}$, with $\hat{I}$ the inversion of $\bm{k}$.
Preferably, we introduce
$
	\tilde{\LL}_a=e^{-i\frac{k_i}{2}}\LL_a,
$
which commute with $\T$ and satisfy the RCA:
\begin{equation}\label{k_L_Cliff}
	\{\tilde{\LL}_a, \tilde{\LL}_b\}=2\delta_{ab}.
\end{equation}
In the following, we shall  analyze first the symmetry algebra at a generic $k$ point and then the high-symmetry points of the BZ.

\begin{table}
    \begin{tabular}{C{1.0cm}|C{1cm}C{1.4cm}C{2.1cm}C{1.2cm}C{1cm}}
		\hline\hline
		$d$ & $C^{p,q}$ & $\overline{p-q}$ & MA & $K$ & $\mathrm{dim}_{\mathbb{C}}$ \\
		\hline
		1 & $C^{1,2}$ & 7  & $\mathbb{R}(2)\oplus\mathbb{R}(2)$  & $\mathbb{Z}\oplus \mathbb{Z}$ & 1  \\
		2 & $C^{2,2}$ & 0 & $\mathbb{R}(4)$ & $\mathbb{Z}$ & 2  \\
		3 & $C^{5,0}$ & 5 & $\mathbb{C}(4)$ & $\mathbb{Z}$ & 4 \\
		4 & $C^{0,6}$ & 2 & $\mathbb{H}(4)$ & $\mathbb{Z}$ & 8  \\
		5 &  $C^{5,2}$ & 3 & $\mathbb{H}(4)\oplus \mathbb{H}(4)$ & $\mathbb{Z}\oplus \mathbb{Z}$ & 8  \\
		6 & $C^{2,6}$ & 4 & $\mathbb{H}(8)$ & $\mathbb{Z}$ & 16\\
		7 & $C^{9,0}$ & 1 & $\mathbb{C}(16)$ & $\mathbb{Z}$ & 16 \\
		8 & $C^{0,10}$ & 6 & $\mathbb{R}(32)$ & $\mathbb{Z}$ & 16\\		\hline\hline
	\end{tabular}
	\caption{RCAs at generic $k$ points with $d=1,\cdots, 8$. $\overline{p-q}$ is the remainder of $p-q\mod 8$. MA stands for matrix algebra. Each $\Z$ component of $K$ corresponds to an IRREP, whose complex dimension is denoted by $\mathrm{dim}_{\mathbb{C}}$. \label{tab:Clifford}}
\end{table}

\begin{table*}
	\begin{tabular}{|c|cccccccc|}
		\hline
                \diagbox{$s$}{$d$}&1&2&3&4&5&6&7&8\\
		\hline
		0& $(\mathbb{R}(2)^{2},1)$ &$(\mathbb{R}(4),2)$&$(\mathbb{C}(4),4)$&$(\mathbb{H}(4)^{32},8)$&$(\mathbb{H}(4)^{64},8)$&$(\mathbb{H}(8)^{64},16)$&$(\mathbb{C}(16)^{ 128},16)$&$(\mathbb{R}(32)^{256},16)$\\
		1& $(\mathbb{R}(4),2)$ &$(\mathbb{R}(4)^4,2)$&$(\mathbb{R}(8)^4,4)$&$(\mathbb{C}(8)^{8},8)$&$(\mathbb{H}(8)^{16},16)$&$(\mathbb{H}(8)^{64},16)$&$(\mathbb{H}(16)^{64},32)$&$(\mathbb{C}(32)^{128},32)$\\
		2& &$(\mathbb{R}(8),4)$&$(\mathbb{R}(8)^4,4)$&$(\mathbb{R}(16)^{4},8)$&$(\mathbb{C}(16)^{8},16)$&$(\mathbb{H}(16)^{16},32)$&$(\mathbb{H}(16)^{64},32)$&$(\mathbb{H}(32)^{64},64)$\\
		3& &&$(\mathbb{R}(16),8)$&$(\mathbb{R}(16)^{8},8)$&$(\mathbb{R}(32)^{4},16)$&$(\mathbb{C}(32)^{8},32)$&$(\mathbb{H}(32)^{16},64)$&$(\mathbb{H}(32)^{64},64)$\\
		4& &&&$(\mathbb{R}(32),16)$&$(\mathbb{R}(32)^{4},16)$&$(\mathbb{R}(64)^{4},32)$&$(\mathbb{C}(64)^{8},64)$&$(\mathbb{H}(64)^{32},128)$\\
		5& &&&&$(\mathbb{R}(64),32)$&$(\mathbb{R}(64)^{4},32)$&$(\mathbb{R}(128)^{4},64)$&$(\mathbb{C}(128)^{16},128)$\\
		6& &&&&&$(\mathbb{R}(128),64)$&$(\mathbb{R}(128)^{4},64)$&$(\mathbb{R}(256)^{8},128)$\\
		7& &&&&&&$(\mathbb{R}(256),128)$&$(\mathbb{R}(256)^{8},128)$\\
		8& &&&&&&&$(\mathbb{R}(512),256)$\\
		\hline
	\end{tabular}
	\caption{RCAs at high-symmetry points $K^{s,d-s}$. Each entry presents $(\mathrm{LA}(s,d),\mathrm{dim}_{\mathbb{C}}(s,d))$. The exponent in each $\mathrm{LA}(s,d)$ indicates the number of IRREPs. \label{tab:Little_Algebra} }
\end{table*}

\section{RCA at generic points}

Consider a generic $\bm k$ in the BZ. The little algebra, namely the symmetries leaving $\bm{k}$ invariant, consists of $\P T$ and $\L_a$.
From \eqref{P_L_Alge} and \eqref{T_ML_Alg}, their operators in momentum space satisfy
\begin{equation}\label{k_PT_Alge}
	\PP\T \tilde{\LL}_a=(-1)^{d+1} \tilde{\LL}_a \PP\T.
\end{equation}

For a given dimension $d$, the operators $\PP\T$, $\LL_a$ and $i$ can be recombined into a RCA. Note that because $\PP\T$ is an anti-unitary operator, the imaginary unit $i$ should also be treated as an operator.
From \eqref{k_PT_Alge}, there are two different cases: $d$ is even or odd.

If $d=2n$ with $n=1,2,\cdots$, the $d+2$ generators are given by
\begin{equation}\label{even_generators}
	\PP\T, \quad i\PP\T, \quad  \tilde{\LL}_a,
\end{equation}
which anticommute with each other. Their squares satisfy
\begin{equation}\label{Square_even}
	(\PP\T)^2=(i\PP\T)^2=(-1)^n,\quad \tilde{\LL}_a^2=1.
\end{equation}

If $d=2n+1$, the anti-commuting generators are
\begin{equation}\label{odd_generators}
	\PP\T, \quad i\PP\T, \quad  i\tilde{\LL}_a,
\end{equation}
with squares
\begin{equation}\label{Square_odd}
	(\PP\T)^2=(i\PP\T)^2=(-1)^n,\quad (i\tilde{\LL}_a)^2=-1.
\end{equation}

It is clear that a RCA structure is unveiled for both cases. For instance, when $d=2$, $\PP\T$ and $i\PP\T$ have negative squares and $\tilde{\LL}_1$ and $\tilde{\LL}_2$ have positive squares, which correspond to the RCA  $C^{2,2}$. Other dimensions can be similarly analyzed.
The results for $d=1,\cdots,8$ are tabulated in Table~\ref{tab:Clifford}. Each RCA is isomorphic to a $(n\times n)$ matrix algebra $\mathbb{K}(n)$ or $\mathbb{K}(n)\oplus \mathbb{K}(n)$ over the ground field $\mathbb{K}=\mathbb{R},\mathbb{C}$, or $\mathbb{H}$, namely real, complex, or quaternion numbers~\cite{Moore_note_2020}. As listed in
Table~\ref{tab:Clifford}, a RCA has at most two IRREPs, with the same rank $\mathrm{dim}_{\mathbb{C}}$. This $\mathrm{dim}_{\mathbb{C}}$ dictates the band degeneracy at the generic $k$ point. More information for IRREPs of RCAs can be found in the Appendix B~\cite{Supp}.

Under the Morita equivalence, there exist only eight independent RCAs determined by $(p-q) \mod 8$. Intriguingly, as shown in Table~\ref{tab:Clifford}, the first eight dimensions exactly exhaust \emph{all} the eight equivalence classes.
Furthermore, all higher dimensions $d>8$ can be reduced to the eight elementary cases. This is because increasing $d$ by $8$ corresponds to rising $p$ ($q$) by $8$ if $d$ is odd (even). Since RCAs satisfy the relation~\cite{atiyah1964clifford,Moore_note_2020}
\begin{equation}\label{Eight_Periodicity}
	C^{p+8,q}\cong C^{p,q+8}\cong \mathbb{R}(16)\otimes C^{p,q},
\end{equation}
the RCAs for $(d+8)$D and $d$D are Morita equivalent. This confirms the eightfold periodicity as we claimed at the  beginning.
From \eqref{Eight_Periodicity}, we see that
\begin{equation}
  \mathrm{dim}_{\mathbb{C}}(d+8)=8\mathrm{dim}_{\mathbb{C}}(d).
\end{equation}
Thus, the band degeneracy for all dimensions can be derived from $d=1,\cdots, 8$.

\section{RCA at high-symmetry points}

For a rectangular lattice, the coordinate components of any high-symmetry point $K$ are either 0 or $\pi$. At these points, there are additional symmetries $\M_a$ and $T$ that need to be considered.
We first note that regarding the symmetry algebra, it is sufficient to analyze only the subset of points
\begin{equation}
	K^{s,r}=(\pi,\cdots,\pi,0,\cdots,0),
\end{equation}
since the little algebra depends only on the number of $\pi$'s and $0$'s in the coordinate. Here, $s$ ($r$) denotes the number of $\pi$ ($0$), and $s+r=d$.

For $K^{s,r}$, $\LL_a$ and $\MM_a$, respectively, satisfy the RCAs:
\begin{equation}
	\{\LL_a,\LL_b\}=2\eta_{ab},\quad \{\MM_a,\MM_b\}=2\delta_{ab},
\end{equation}
Since $\LL_a^2=e^{iK_a^{s,r}}$, $\eta_{ab}$ is the diagonal matrix with first $s$ diagonal elements being $-1$ and the others being $1$.  
However, their mutual algebraic relations are quite nontrivial:
\begin{equation}
	\MM_a \LL_a \MM_a=\LL_a^{-1}=\eta_{aa} \LL_a,\quad \{\MM_a,\LL_b\}=0
\end{equation}
with $a\ne b$.
Hence, the first $s$ $\MM$'s anticommute with all $\LL$'s. This means
\begin{equation}\label{Cliff_s_d}
	\MM_1,\cdots, \MM_s, \LL_1,\cdots,\LL_d
\end{equation}
anticommute with each other. They form a RCA $C^{s,d}$, since the first $s$ $\LL$'s square to $-1$, and the remaining $\LL$'s and all $\MM$'s square to $+1$.

Meanwhile, each $\MM_{a}$ with $s<a\le d$ anticommutes with almost all $\LL$'s and $\MM$'s except $\LL_a$. The algebraic structure can be dissolved by turning to the combinations:
\begin{equation}\label{Cliff_C01}
	\MM_{s+1}\LL_{s+1},\cdots, \MM_{d}\LL_{d},
\end{equation}
which commute with each other and with all operators in \eqref{Cliff_s_d}. Since $(\MM_{a}\LL_a)^2=1$ with $s<a\le d$, each $\MM_{a}\LL_a$ generates a RCA $C^{0,1}$.

From \eqref{Cliff_s_d} and \eqref{Cliff_C01}, we conclude that the little algebra at $K^{s,r}$ is given by
\begin{equation}\label{Little_Alg_K}
\mathrm{LA}(s,d)\cong C^{s,d}\otimes (C^{0,1})^{\otimes(d-s)}\otimes C^{1,1}.
\end{equation}
The last component $C^{1,1}$ is the Clifford algebra generated by $\T$ and $i$. The algebra \eqref{Little_Alg_K} can be simplified as
\begin{equation}
	\mathrm{LA}(s,d)\cong [\mathbb{R}(2^{s+1})]^{\oplus 2^{d-s}}\otimes C^{0,d-s}.
\end{equation}
Thus all the IRREPs of $\mathrm{LA}(s,d)$ have the same rank~\cite{atiyah1964clifford,Moore_note_2020}
\begin{equation}
	\mathrm{dim}_{\mathbb{C}}(s,d)=2^{s}\mathrm{dim}_{\mathbb{R}}(M^{0,d-s}).
\end{equation}
We observe that for each dimension $d$, $\mathrm{dim}_{\mathbb{C}}(s,d)$ is monotonically increasing with $s$. The results for $d=1,\cdots,8$ are shown in Table~\ref{tab:Little_Algebra}.

\begin{figure}
	\centering
        \includegraphics[width=0.48\textwidth]{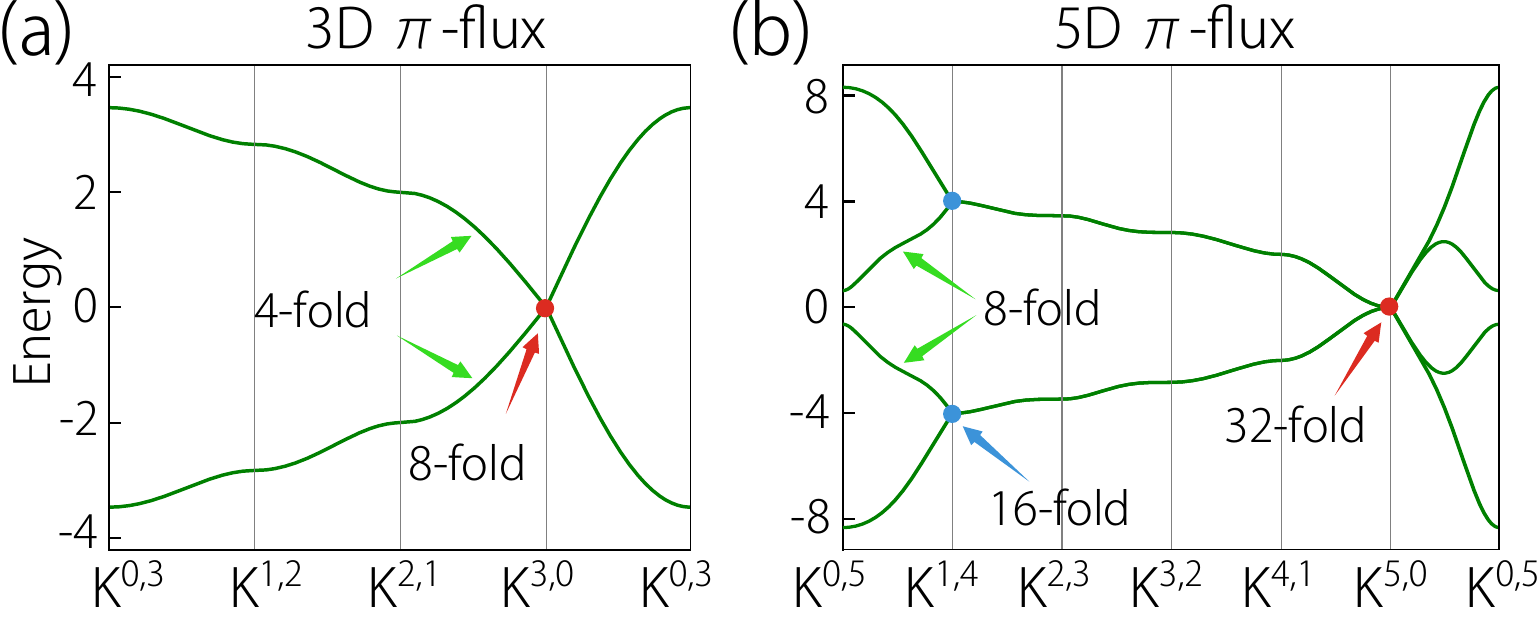}
	\caption{Band structures for $\pi$-flux rectangular lattices. (a) shows a $3$D model and (b) shows a 5D model. Note the large degeneracy
at $K^{d,0}$ point.}
	\label{fig-spectrum35}
\end{figure}

\section{Topological states from dimerization}

The ranks of IRREPs in Table~\ref{tab:Clifford} and \ref{tab:Little_Algebra} are just the corresponding degrees of degeneracy of the band structure. This can be explicitly confirmed by numerical calculations (see Appendix C for more details~\cite{Supp}). The results for $d=3$ and $d=5$ are shown in Fig.~\ref{fig-spectrum35}. The degeneracy degrees for the generic $k$ point and for high-symmetry points agree with Table~\ref{tab:Clifford} and \ref{tab:Little_Algebra}.

\begin{figure}[tb]
    \centering
    \includegraphics[width=0.43\textwidth]{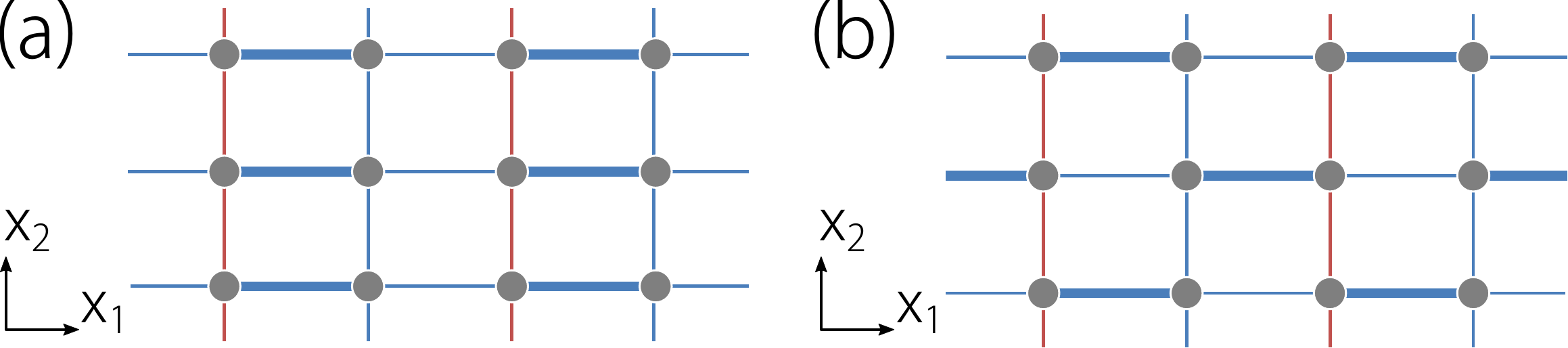}
    \caption{ Two basic dimerization patterns.
        (a) The dimerization is added along the $x_1$ direction and is
        uniform along $x_2$. (b) The dimerization is added along $x_1$ and it alternates along $x_2$.}
    \label{fig-dimerization}
\end{figure}

Band degeneracy is a resource for generating topological phases. Here, we have large degeneracy arising from the RCAs in a simple rectangular lattice. We will consider the strategy of dimerization to drive the system into various topological phases. The two basic dimerization patterns are illustrated in Fig.~\ref{fig-dimerization}.
Below, we exemplify the general mechanism by presenting some fascinating examples.

\begin{figure}
    \centering
    \includegraphics[width=0.45\textwidth]{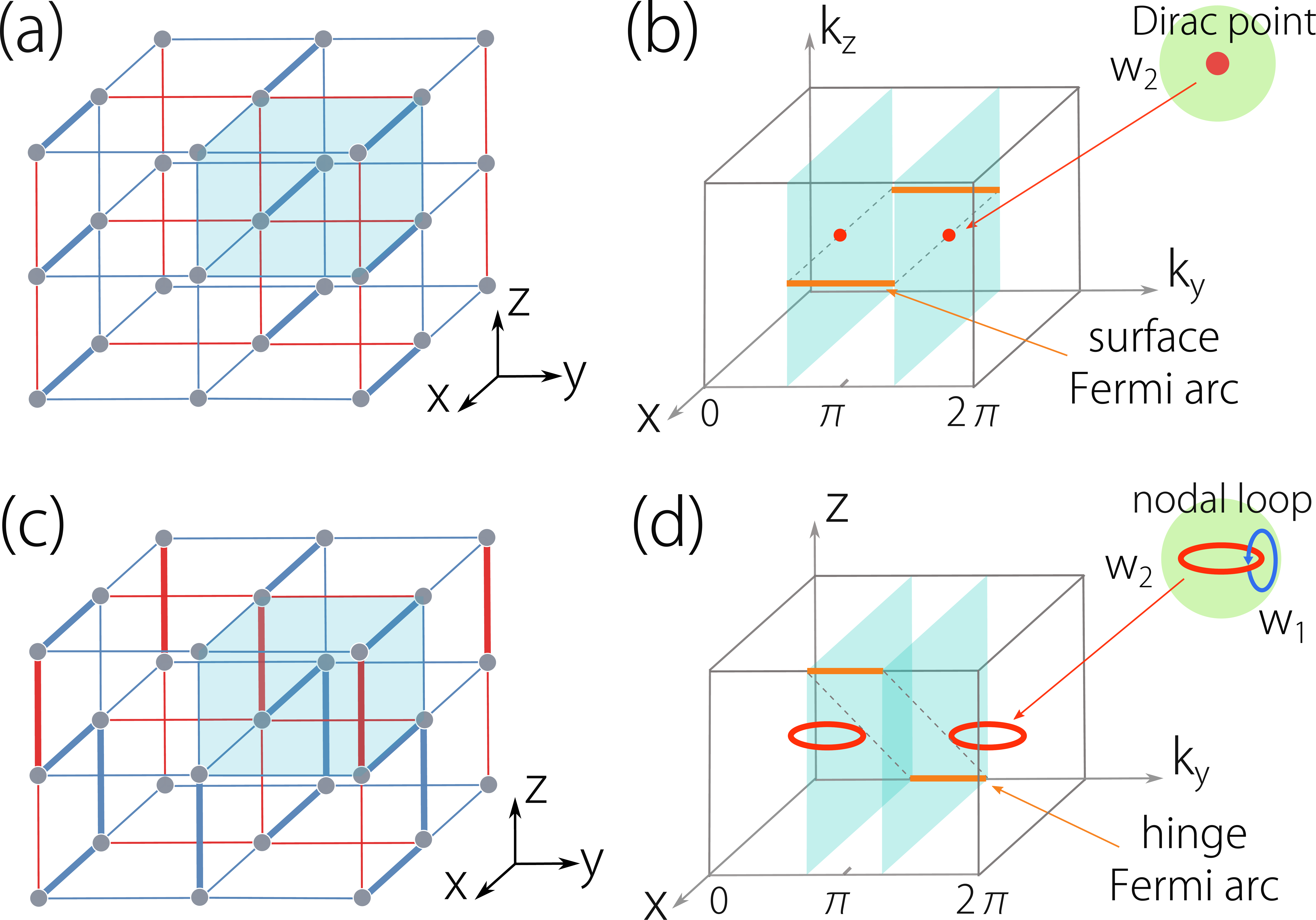}
    \caption{(a) $3$D $\pi$-flux lattice with a dimerization added along $x$. (b) The nodal point at $K^{3,0}$ splits into
        two real Dirac points, with a pair of surface Fermi arcs. (c) Further adding to (a) a dimerization along the $z$ direction. (d) Each Dirac point in (b) transforms to a nodal loop, with a pair of hinge Fermi arcs. The insets in (b) and (d) indicate the topological charges of the nodal structures.}
    \label{fig-3d}
\end{figure}

Our first example is the $3$D higher-order nodal-loop semimetal state
with twofold Stiefel-Whitney topological charges. Starting from the $3$D $\pi$-flux lattice, we first add dimerization along $x$ which alternates along $y$ but is uniform along $z$ [Fig.~\ref{fig-3d}(a)], which breaks $\L_x$ and $\L_y$. Then, the $8$-fold nodal point at $K^{3,0}$ is split into two $4$-fold real Dirac points, protected by $\L_x\L_z\P T$, as illustrated in Fig.~\ref{fig-3d}(b). Although $\P$ is broken, the off-centered inversion symmetry $\L_x\L_z\P$ is preserved, with $(\L_x\L_z\P)^2=1$ [see Eq.~\eqref{P_L_Alge}]. The real Dirac points feature $2$D Stiefel-Whitney topological charge $w_2=1$~\cite{zhao_pt-symmetric_2017}. Further adding dimerization along $z$ which alternates along both $x$ and $y$ [Fig.~\ref{fig-3d}(c)] will spread each Dirac point into a nodal loop [Fig.~\ref{fig-3d}(d)]. Each nodal loop not only inherits $w_2$, but has an additional $1$D Stiefel-Whitney charge $w_1$, namely the quantized $\pi$ Berry phase.
The $2$D charge $w_2$ leads to hinge Fermi arcs on a pair of inversion-related edges [Fig.~\ref{fig-3d}(d)], and the 1D charge $w_1$ will lead to drumhead surface modes.

\begin{figure}
    \centering
    \includegraphics[width=0.43\textwidth]{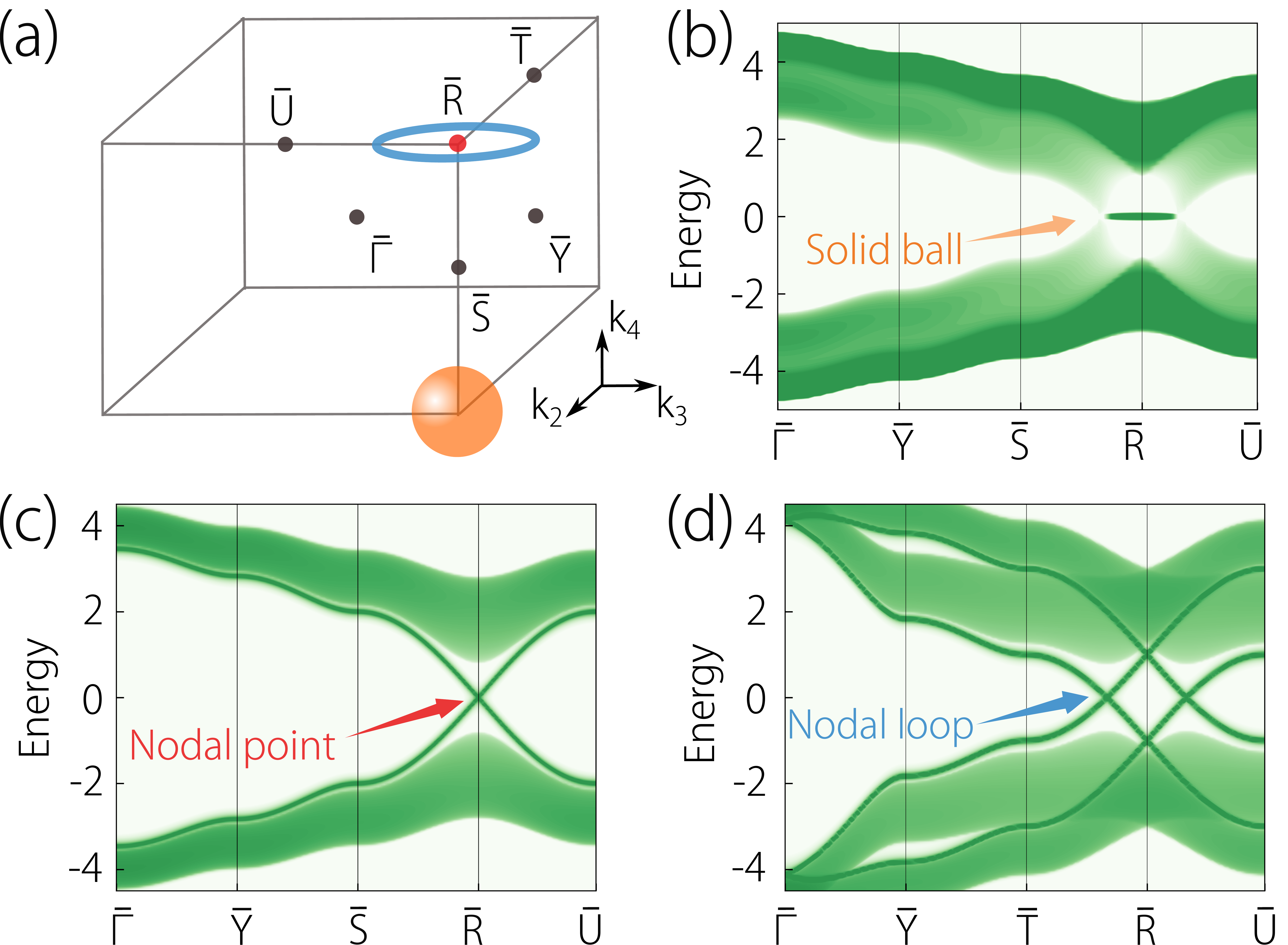}
    \caption{(a) $3$D boundary BZ for the $4$D model. The boundary zero-modes with the shape of (orange) solid ball, (red) 8-fold nodal point, and (blue) 4-fold nodal loop are indicated, respectively correspond to the boundary spectra in (b), (c), and (d).
    The dimerization patterns that generate these novel states are discussed in the main text.
        }
    \label{fig-4d}
\end{figure}

As another example, we consider the $4$D $\pi$-flux lattice. Our first attempt is to add a dimerization along $x_1$ which alternates in all other three directions. This breaks all unit translations $\L_a$.
We find that the original $16$-fold nodal point at $K^{4,0}$ evolves into a novel $8$-fold nodal surface $S^2$, i.e., the nodal manifold is a 2-sphere in the $k_2$-$k_3$-$k_4$ sub-BZ for $k_1=\pi$. At the boundary normal to $x_1$,  we find  zero modes occupying a solid-ball region in the $3$D boundary BZ [Fig.~\ref{fig-4d}(a,b)].

Our second attempt is to add a dimerization along $x_1$ but the pattern is uniform along all other three directions. Then, the bulk spectrum will be gapped out and the system becomes a $4$D M\"obius topological insulator. On the $3$D boundary normal to $x_1$, there exists an $8$-fold surface nodal point at the $\bar{R}$ point [Fig.~\ref{fig-4d}(a,c)]. Since the eigenvalues of the translations $\L_a$ with $a=2,3,4$ all have period $4\pi$, the boundary bands exhibit a M\"obius-twist in terms of the $\L_a$ eigenvalues. If we further add a dimerization along $x_4$ which is
uniform along $x_1$ but alternates along $x_2$ and $x_3$, then all the $\L_a$'s will be broken. The bulk remains insulating and the $8$-fold surface nodal point will spread into a 4-fold surface nodal loop around $\bar{R}$ [Fig.~\ref{fig-4d}(a,d)].

\section{Discussion}
We have unveiled that all RCA equivalence classes can be elegantly realized in a very simple lattice setup as projective symmetry algebras. The realization exhibits the eightfold Bott periodicity with spatial dimension of the lattice. The IRREPs of RCAs directly determine the degeneracies in the band structures of the lattice. The findings are not only of fundamental interest, but also offers a new route to achieve unprecedented topological phases. For instance, the higher-order nodal-loop semimetal state was initially proposed using a Dirac model~\cite{wang_boundary_2020}, which is difficult to find an experimental realization. Here, via RCAs and dimerization strategy, we achieve this state in a simple rectangular lattice, which can be readily
realized, e.g., in acoustic crystals~\cite{li_acoustic_2021,xue_projectively_2021}. The construction here also provide a general approach to topological phases in higher dimensions, as exemplified by our $4$D examples.

\begin{acknowledgements}
  The authors thank S. Wu and D. L. Deng for helpful discussions. This work is supported by  National Natural Science Foundation of China (Grants No. 12174181, No. 12161160315, and No. 11874201), and the Singapore MOE AcRF Tier 2 (T2EP50220-0026). We acknowledge computational support from Texas Advanced Computing Center.
\end{acknowledgements}

\appendix
\setcounter{figure}{0}
\renewcommand{\thefigure}{S\arabic{figure}}
\section{Derivation of projective algebras}
\setcounter{equation}{0}
\renewcommand{\theequation}{A\arabic{equation}}
\begin{figure}[h]
    \centering
    \includegraphics[width=0.41\textwidth]{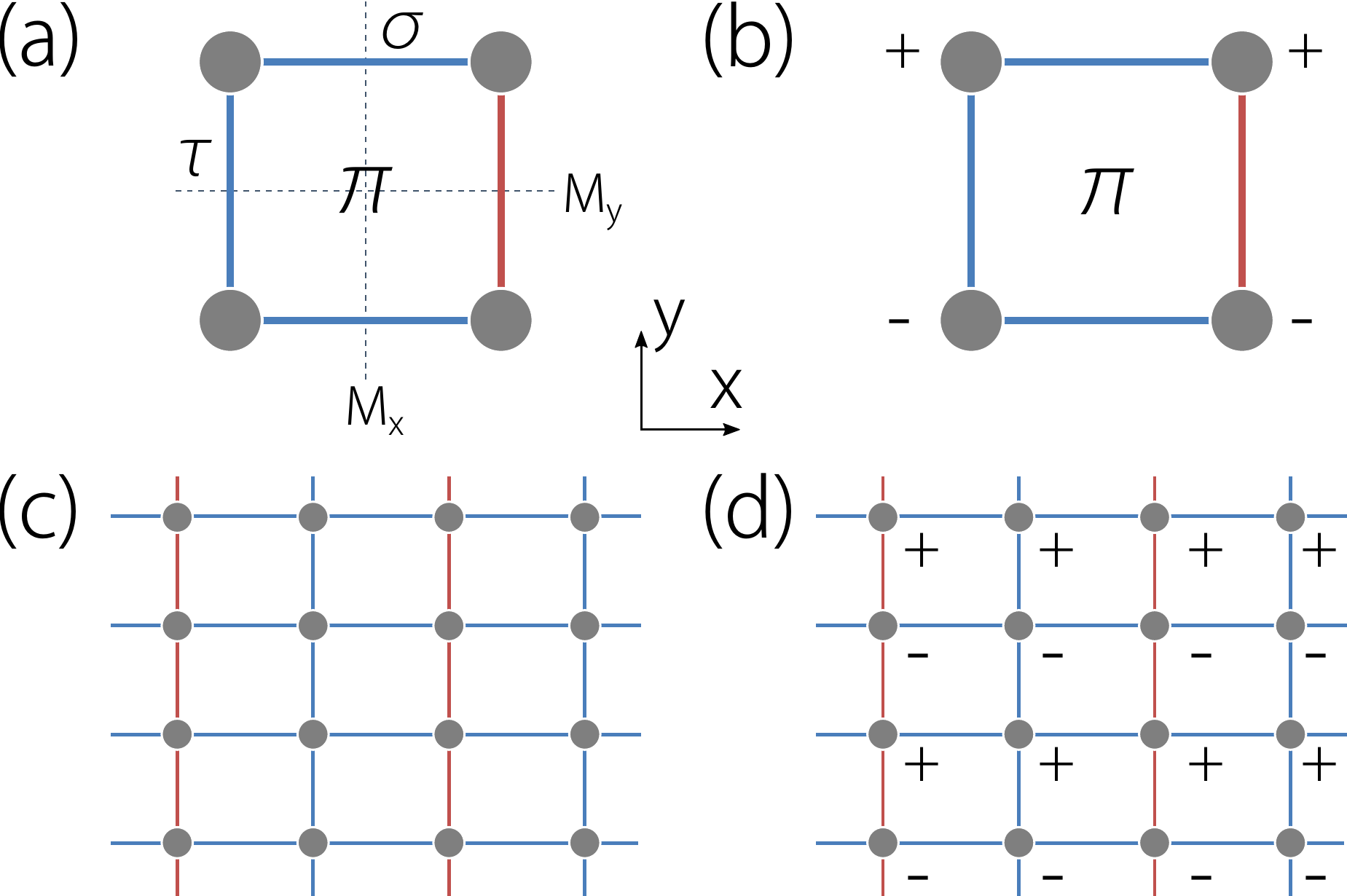}
    \caption{(a) The gauge configuration in a rectangular unit cell. (b) The gauge transformation $G=\sigma_0\otimes \tau_3$ in the
        cell. (c) One gauge configuration for the $2$D $\pi$-flux  lattice. (d) Gauge transformation in the $2$D lattice.}
    \label{fig-S1}
\end{figure}
In this appendix, we derive the projective algebras of $\L_a$ and $\M_a$ in the main text. On a lattice with $\pi$ fluxes, a spatial symmetry $S$ may preserve the flux configuration, but may not preserve the gauge connection configuration, namely the configuration of hopping phases. Then, after the transformation of $S$, a gauge transformation $\mathsf{G}_{S}$ should be applied in order to restore the original gauge connection configuration. Thus, the \emph{proper} symmetry operator is a combination~\cite{zhao_mathbbz_2-projective_2020}:
\begin{equation}
	\mathsf{S}=\mathsf{G}_S S.
\end{equation}
Because of the accompanied gauge transformations $\mathsf{G}_S$, the original group algebra of spatial symmetries will be modified to be a projective group algebra.

To understand the projective algebra $\{\M_a,\M_b\}=2\delta_{ab}$ in the main text, it suffices to consider a single rectangular plaquette with flux $\pi$. As illustrated in Fig.~\ref{fig-S1}(c), the spatial mirror operators are given by
\begin{equation}
	M_x=\sigma_1\otimes\tau_0,\quad M_y=\sigma_0\otimes \tau_1.
\end{equation}
After the mirror reflections, the gauge connection configuration is generally changed. Particularly, for the gauge choice in Fig.~\ref{fig-S1}(c), the required gauge transformations are given by
\begin{equation}
	\mathsf{G}_{M_x}=\sigma_0\otimes\tau_3,\quad \mathsf{G}_{M_y}=\sigma_0\otimes\tau_0.
\end{equation}
Here, $\mathsf{G}_{M_y}$ is the identity operator because the gauge choice preserves $M_y$.
Then, the proper mirror operators are given by
\begin{equation}
	\M_x=\mathsf{G}_{M_x}M_x=\sigma_1\otimes\tau_3,\quad \M_y=M_y=\sigma_0\otimes\tau_1.
\end{equation}
It is now evident that
\begin{equation}
	\{\M_x,\M_y\}=0,\quad \M_x^2=\M_y^2=1.
\end{equation}

We then repeat the plaquette infinitely along the $x$ and $y$ directions to form a $2$D rectangular lattice. Manifestly, the gauge connection configuration is invariant under $L_y$. Hence, $\L_y=L_y$. However, $L_x$ changes the gauge connection configuration, and therefore, $L_x$ should be accompanied with a gauge transformation $\mathsf{G}_{L_x}$, as illustrated in Fig.~\ref{fig-S1}(d). It is observed that $\mathsf{G}_{L_x}$ is odd under $L_y$, i.e.,
\begin{equation}
	L_y \mathsf{G}_{L_x}L_y^{-1}=-\mathsf{G}_{L_x}.
\end{equation}
Thus, for
\begin{equation}
	\L_x=\mathsf{G}_{L_x}L_x, \quad \L_y=L_y,
\end{equation}
we have
\begin{equation}
	\{\L_x,\L_y\}=0.
\end{equation}

On the rectangular lattice, it is obvious that
\begin{equation}
	\mathsf{G}_{M_x}=\mathsf{G}_{L_x},
\end{equation}
and therefore,
\begin{equation}
	\M_x=\mathsf{G}_{L_x}M_x, \quad \M_y=M_y.
\end{equation}
Moreover, $L_x$ and $M_x$ both preserve $\mathsf{G}_{L_x}$, i.e.,
\begin{equation}
	L_x\mathsf{G}_{L_x} L_x^{-1}=\mathsf{G}_{L_x},\quad M_x \mathsf{G}_{L_x} M_x=\mathsf{G}_{L_x},
\end{equation}
But $M_y$ inverses $\mathsf{G}_{L_x}$, i.e.,
\begin{equation}
	M_y\mathsf{G}_{L_x}M_y=-\mathsf{G}_{L_x}.
\end{equation}
Thus, we conclude that
\begin{equation}
	\M_x \L_x \M_x=\L_x^{-1},\quad \{\M_x, \L_y\}=0
\end{equation}
\begin{equation}
	\M_y \L_y \M_y=\L_y^{-1},\quad \{\M_y, \L_x\}=0.
\end{equation}


\section{IRREPs of real Clifford algebras}
\setcounter{equation}{0}
\renewcommand{\theequation}{B\arabic{equation}}
The RCA have an elegant representation theory. In this section, we briefly collect basic results on the IRREPs of RCA.

$C^{p,q}$ has two IRREPs with the same dimension if $p-q=3$ or $7 \mod 8$. Otherwise, $C^{p,q}$ has a unique IRREP. Here, `$\mathrm{mod}~8$' essentially comes from~\cite{atiyah1964clifford,Moore_note_2020}
\begin{equation}\label{Eight_Periodicity1}
	C^{p+8,q}\cong C^{p,q+8}\cong \mathbb{R}(16)\otimes C^{p,q}.
\end{equation}

Each real Clifford algebra is isomorphic to  $\mathbb{K}(2^\ell)\oplus\mathbb{K}(2^\ell)$ with $\mathbb{K}=\mathbb{R}$ or $\mathbb{H}$ if $p-q=3$ or $7 \mod 8$, or otherwise is isomorphic to $\mathbb{K}(2^\ell)$ with $\mathbb{K}=\mathbb{R}$, $\mathbb{C}$ or $\mathbb{H}$. Here, $\mathbb{R}, \mathbb{C}, \mathbb{H}$ denote real, complex, and quaternion numbers, respectively. For $d=1,2,\cdots,8$, the matrix algebras are worked out in Table I in the main text. The complex dimension $\mathrm{dim}_\mathbb{C}$ of IRREPs is given by $2^{\ell-1}$, $2^{\ell}$, $2^{\ell+1}$ for $\mathbb{K}=\mathbb{R}$, $\mathbb{C}$ and $\mathbb{H}$, respectively. With the fact that $\mathbb{R}(n)\otimes \mathbb{K}(m)\cong \mathbb{K}(nm)$, we can derive $\mathrm{dim}_{\mathbb{C}}$ for all higher dimensions using \eqref{Eight_Periodicity1}. The degeneracy for a generic $\bm{k}$ in $d$D BZ is given by
\begin{equation}
	\mathrm{dim}_{\mathbb{C}}(d)=16^k\mathrm{dim}_{\mathbb{C}}(d^r).
\end{equation}
Here, $d=8k+d^r$ with $0<d^r<8$, and $\mathrm{dim}_{\mathbb{C}}(d^r)$ can be found in Table I.

\section{Lattice models}
\setcounter{equation}{0}
\renewcommand{\theequation}{C\arabic{equation}}
As mentioned above, due to the AB effect, any two different translation operators anti-commute
with each other in the $d$D system. Typically, we specify the lattice with $\L_a^2$
and $\L_b$ since $\comm{\L_a^2}{\L_b}=0$ in the corresponding directions. In this
case, it is straightforward to have the relation
$\L_b \L_a \psi(\bm k)=e^{i(k_b+\pi)}\L_a\psi(\bm k)$
if one assume $\psi(\bm k)$ be an eigenstate of $\L_b$, namely $\L_b\psi(\bm k)=e^{ik_b}\psi(\bm k)$. Thus, although $\L_a\psi(\bm k)$
has the same energy with $\psi(\bm k)$, the momentum is $(k_a,k_b+\pi)$, showing the spectrum having a $\pi$-periodicity along $k_b$.
This leads us to fold the BZ, or double the unit cell. Generalized to $n$-dimension system, the final BZ should be specified
by $(\L_1^2,\L_2^2,\dots)$. Therefore, the unit cell we considered has $2^n$ atoms under the $\mathbb Z_2$ field.
To construct the tight-binding model in $d$D, assuming we have $\pi$-flux system in the $(d-1)$-dimensional system $H_{d-1}$,
the stacking of $H_{d-1}$ along the $d$-th direction leads to $H_{d-1}\otimes \sigma_0$.
Note that the simple stacking does not introduce the desired flux we need in the planes along the additive direction.
Heuristically, if we engineer the hopping along $d$-th direction with opposite signs between the nearest sites,
all of the small rectangular plaquettes must have a $\pi$ flux in all of the planes, that is
\begin{eqnarray}
	H_d= H_{d-1}\otimes \sigma_0+t(1+\cos k_d)\Gamma_{3,\dots,3,1}+t \sin k_d \Gamma_{3,\dots,3,2},
	\nonumber\\
	\label{eq-Hdm1Hd}
\end{eqnarray}
where $\Gamma_{i,j,k,\dots}=\sigma_i\otimes \sigma_j\otimes \sigma_k\otimes \dots$, with the $n^\mathrm{th}$ Pauli matrix acting on the $n^\mathrm{th}$ direction.
In this gauge configuration, the translation operator along one direction can be represented by an off-diagonal matrix
\begin{eqnarray}
	T_m=\mqty[0 & 1 \\ e^{ik_m} & 0].
	\label{eq-Tioperator}
\end{eqnarray}
The phase factor $e^{ik_m}$ comes from the translation into the next cell. For an Abelian case, no gauge potential exists, translation
operation along any directions are not entangled, the translations have the form of $L_m=\sigma_0^{\otimes(m-1)}\otimes T_m\otimes \sigma_0^{\otimes(d-m)}$.
Now, one must fix the particular gauge configuration to obtain our proper translation operators, which we have discussed in Appendix A for $2$D.
As stated above, the hopping signs in the higher directions have a staggered sign pattern along the lower dimension sites, then it is straightforward to write down
\begin{eqnarray}
	\L_m= \sigma_0^{\otimes (m-1)} \otimes T_m \otimes \sigma_3^{\otimes (d-m)},
	\label{eq-Lmoperator}
\end{eqnarray}
which immediately yields the anti-commutation relation $\acomm{\L_m}{\L_n}=2\delta_{mn}\L_m^2$.
Using the similar method, the mirror operators are represented by
\begin{eqnarray}
	\M_m=  \mathsf I_m
	\sigma_0^{\otimes (m-1)}\otimes \sigma_1 \otimes \sigma_3^{\otimes (d-m)},
	\label{eq-mirrorop}
\end{eqnarray}
where $\mathsf I_m=(k_m\to -k_m)$ inverts $k_m$ due to the mirror operation. Finally, the product of the mirror operators
gives the inversion operator,
\begin{eqnarray}
	\P= \mathsf I_{\bm k}
	\sigma_1\otimes i\sigma_2 \otimes \dots \otimes (\sigma_3^{i-1} \sigma_1) \otimes \dots,
	\label{eq-inversion operator}
\end{eqnarray}
whose square strongly depends on the dimensionality and satisfies Eq.~(5) in the main text. With $\L_i$ and $\P\T$, one can verify the algebraic relations
and degeneracy at generic points in Table I and II.
For $d=3$, an $8$-fold point is enforced at point $K^{3,0}$, and it deforms into two isolated and chiral bands with $4$-fold
degeneracy deviating from $K^{3,0}$. We remark that the IRREPs of the RCAs only guarantee the minimal degeneracy
of the points in BZ. Other space groups or symmetries may introduces extra protection, and higher degeneracy may emerge. For $d=5$,
the generic $\bm k$ of $H_5$ in Eq.~\eqref{eq-Hdm1Hd} shows $16$-fold degeneracy, while our prediction gives $8$. Actually, due to
the redundant dimension in space, an extra hopping perturbation, between the diagonal site of the $5$D cubic is allowed,
\begin{eqnarray}
	H_5'= t' P_1\otimes Q_2\otimes P_3\otimes Q_4\otimes P_5,
	\label{eq-H5p}
\end{eqnarray}
where $P_i=(1+\cos k_i)\sigma_1+\sin k_i \sigma_2$ and $Q_i=\sin k_i \sigma_2-(1+\cos k_i)\sigma_2$.
As we can see, $Q_i$ anti-commutes with time-reversal operator, therefore similar terms are forbidden in $d=2,3$.
In $d=4$, it does not modify the degeneracy. However, in $d=5$ this term has its effect, as shown in Fig. 2(b),
the degeneracy at generic $\bm k$ point is $8$; and at point $K^{s,r}$, $0<s<5$, the degeneracy is $16$ since this
term vanishes at any high-symmetry points; finally, all the bands join at $K^{5,0}$.

\bibliographystyle{apsrev4-2}
\bibliography{Eight_bot}

\end{document}